\documentstyle [10pt,fleqn]{article}
\title{ Low Dimensional Dynamics \\
in a Pulsating Star \footnote{
Based on observations made with the NASA/ESA Hubble Space Telescope,
obtained at the Space Telescope Science Institute, which is operated
by the Association of Universities for Research in Astronomy, Inc., under
NASA contract NAS5-26555.}
\\*[1.5cm]}

\author{ G. B. Mindlin $\! ^1$, P. T. Boyd $\! ^2$, J. L. Caminos $ \! ^3$,
J. A. Nu\~nez $\! ^3$\\
{\small\it $\! ^1$ Departamento de F\'{\i}sica, FCEN } \\
{\small\it Ciudad Universitaria, Pab. I, c. p. 1428, Buenos Aires,
Argentina
} \\
{\small\it $\! ^2$ Universities Space Research Association and
 Laboratory for High Energy Astrophysics } \\
{\small\it NASA, Goddard Space Flight Center, Greenbelt,
MD 20771
} \\
{\small\it $\!^3$
 Observatorio Astronomico, FCGLP} \\
{\small\it Universidad de La Plata, c. p. 1900, La Plata, Argentina
} \\
}
\date{}
\begin{document}

\maketitle
\vspace*{1.5 cm}
\begin{abstract}

We report the discovery of a low dimensional dynamical system
in a 5.5 hr Hubble Space Telescope High Speed Photometer
observation of a rapidly oscillating
star.
The topological description of the phase space orbits
is given, as well as a dynamical model which
describes the results.  This model should motivate theorists
of stellar pulsations to search for a three-dimensional system with the
same topological structure to describe
 the mechanisms for pulsation.
The equations are compatible with recently proposed nonlinear mode interaction
 models.

\end{abstract}
\vfill
\eject

 The rapidly oscillating peculiar A stars (hereafter $roAp$ stars)
were identified as a class of pulsating single stars in 1978 by Kurtz
(cf, \cite{kurz90}).  They exhibit nearly regular fluctuations, on the
order of
milli-magnitudes, on time scales of 3 to 25 minutes, believed to be
low order, high overtone, non-radial p-mode oscillations.  The oscillating
material is localized near the magnetic poles; the strength of the oscillations
is modulated by the rotation of the star since the rotation and magnetic axes
of the body are typically not aligned.  Many $roAp$ stars have a rich spectrum
of modes of oscillation, and in these cases there are commonly additional
modulations present which are not so well understood, but are believed to
be related to the non-linear interaction of the active modes \cite{kre91},
\cite{mat87},\cite{mart93}.

 The bulk of research involving these stars concerns the determination
of stellar parameters
 \cite{cox80}.  It is possible to determine the angle
between the rotation and magnetic axes, as well as the angle between the
rotation axis and the observer, through measurement of the frequency
splitting present in very long, often discontinuous, data sets.  This analysis
involves generating a single power spectrum from a large time series.
 The {\em evolution} of the frequencies present has not gained as much
attention, but is at least as interesting, because it can give insight into
the physical mechanism(s) responsible for the oscillations themselves.

The High Speed Photometer (HSP) was a first generation
instrument aboard the Hubble Space Telescope, uniquely capable of making high
time-resolution observations (up to 10,000 Hz) in the ultraviolet-to-visible
regions of the spectrum.  Due to its location in space, it was also
the only photometer free from the effects of atmospheric scintillation.

The
HSP observed the $roAp$
star HD 60435 as a calibration target for 5.5 hours 1991 August 22,
 since the rapid fluctuations
are only a small fraction of its otherwise constant magnitude.
The data were taken with 82.4 millisecond time resolution through a
broad-band filter with peak transmission at 2400 Angstroms and full
width at half-maximum of 550 Angstroms.
Preliminary analysis showed modulation of the main frequency present (11.6
min) on the time scale of the observation, which was identified
as consistent with the beating of 5 frequencies \cite{tay93}.
Closer analysis revealed that this frequency underwent an abrupt switch from on
to off, inconsistent with beating and consistent with nonlinear behavior
\cite{boyd95}.

Effects due to the telescope itself were eliminated according to the
method described elsewhere \cite{tay93}.
Since we are interested in only the behavior of the
relatively long 11.6 minute period, we eliminated high frequency noise by
the following method.  The data were rebinned to 30 second bins, and a
spline interpolation was used to resample back to the original
number of points.  This faithfully represents the longer period oscillations
(a few minutes or longer) while ignoring random point-to-point fluctuations.
This data is shown if Fig. 1.
Even to the eye, there is a marked difference in the behavior of the first
2/3 of this data set compared to the last 1/3.
In Fig. 2 we display a contour plot of the Gabor transform of this data.
This windowed Fourier analysis method clearly shows the power in the 11.6
min frequency undergo a sudden decrease approximately 200 min into the
observation.  Also evident is a weaker structure with slowly increasing
frequency (near 80 cycles early in the observation, increasing toward 120
cycles at later times).

The basic equations describing the dynamics of a pulsating star have
been discussed thoroughly in the literature. These account for the
conservation of energy, momentum, heat gain and losses, as well as for
the equations of state that relate pressure to density and temperature.
Altogether, these equations describe the competition between radiation
and gravity that is ultimately responsible for the pulsating dynamics.
Even so, the lack of a qualitative theory of partial differential equations,
combined with the difficulty of measuring the appropriate parameters, makes it
difficult
to treat this problem from first principles. From a theoretical standpoint,
simplified models have been proposed which assume either linearizations
around equilibrium states or weakly nonlinear interactions between active
modes. In the latter case, a reduction of the dynamics (from partial
differential equations to a finite set of ordinary differential equations
describing the behavior of the mode amplitudes) makes the problem tractable.
In support of these simplifying assumptions, it has been reported recently that
the dynamics of pulsating star R Scuti is actually a low dimensional one
\cite{buc95}.

\par

Whenever a complicated invariant set coexists with a set of unstable
periodic orbits, the trajectory corresponding to an initial condition within
invariant set will reflect their influence in the
 existence of $p-$close return segments \cite{mind91}.
These are segments of the scalar time
series $x(i), i=1,...,n$ such that $x(i+k) \sim x(i+p+k) $,
for some sequence of values $k=0,1,2,..$. The relationship between the
existence of unstable periodic orbits buried within the invariant set and the
close returns is the following: whenever a state point is near an unstable
periodic orbit, it can evolve along the stable manifold of the orbit until
it gets expelled along the unstable manifold. If the period of the unstable
periodic orbit is reasonably
small, and the orbit is not highly unstable, the evolution in the neighborhood
of the orbit can be long enough to remain in an epsilon neighborhood
of the starting point for at least one cycle of the periodic orbit.
 Selecting pieces of the scalar data which are
good close returns (almost close themselves and continue behaving similarly
for a while), results in a collection of approximations of the unstable  orbits
coexisting
with the complex trajectory under study.
 Periodic orbits can be embedded in
three dimensions with the use of techniques which aim at
reconstructing a phase space for the orbits.  Among these, the
time delay technique is the most widely used. This consists of creating a
ariate environment
from $x(t) \rightarrow (x(t),x(t-\tau),x(t-2\tau))$. Some close returns found
in the HSP observation of HD 60435, after applying the above
embedding, are displayed in Figs. 3a, 4a and 5a.

We propose that the following low dimensional dynamical system is capable
of reproducing the unstable periodic orbits in the observational
time series:

\begin{eqnarray}
X' & = & Y \nonumber\\
Y' & = & \mu (1+\epsilon cos(\phi))  -Y- X^2 + XY \nonumber\\
\phi' & = & \omega ,  \label{eq:prim}
\end{eqnarray}

\noindent
where X and Y are dynamical variables, $\epsilon$ and $\mu$ are
control parameters and $\omega$ is a constant.
In Figs. 3b, 4b and 5b we display different
(embedded) periodic orbits that are
attracting solutions of equations \ref{eq:prim}, for parameter values
listed in the figure captions. Notice the striking similarity between the
observational orbits and the theoretical ones.

To understand why equations \ref{eq:prim} reproduce the observed orbits, we
describe the dynamics of the equations in the absence of forcing
($\epsilon=0$). This set of two nonlinear coupled equations constitutes the
unfolding of a Takens-Bogdanov bifurcation \cite{guck83}.
For negative values of $\mu$, there
are no fixed points. For $0<\mu<\mu_c$, a pair of fixed points arise: a
saddle (at $x_1<0, y_1=0$) and a sink (at $x_2>0, y_2=0$). As $\mu$ is
increased, the $x$ coordinate distance between the two fixed
point increases; at $\mu_c=(\frac{10-\sqrt(96)}{2})^2$, the fixed point
($x_2,y_2$) becomes an attracting focus. At $\mu=1$, this fixed point
undergoes another qualitative change, and a periodic orbit bifurcates from the
fixed point. If the value of $\mu$ is further increased, the distance between
the two fixed points continues to increase and the periodic orbit gets
closer to the fixed point at $(x_1,y_1)$. As the period of this
periodic orbit increases, the time spent close to this fixed point
is a large fraction of the period. This phenomenon is known as critical
slowing down.

The addition of a forcing term changes the dynamics qualitatively. As the
forcing is in the parameter that controls the distance between the two
fixed points, the dynamics manifests itself as oscillations around a
point that is itself positioned at an oscillating coordinate. Moreover, when e
trajectory gets close to the saddle fixed point, the evolution is slower
than when the trajectory is far away from it, explaining the
uneven distribution
of "curls" in the orbits. A rich variety of dynamical structures can be found
in the equations,
from periodic orbits to chaotic trajectories.

There is a quantitative way to
compare these orbits, by means of topology \cite{kauf91}.
These one dimensional curves
can be embedded in three dimensions, and
can therefore be characterized by the
way in which they are knotted. It is possible to associate with each class of
knots (a class of knots defined as the closed curves that can be deformed
into each other) a {\em knot invariant}. This invariant need not characterize
the class uniquely, but if the invariants associated with two knots are
not equal, then the knots cannot be deformed one into the other (in knot
theoretical language, they are not isotopic). A regular isotopy invariant
is the {\em HOMFLY} polynomial (the term regular relates to the kind of
moves allowed in the process of deforming one knot into the other).

To build this invariant for a given knot, it is necessary first to
obtain a {\em diagram} for the knot, {\em i.e.} a two dimensional projection
of the knot in which small segments of the projected curve are deleted in order
to keep information on the overcrossings taking place in the real 3-dimensional
space (see Figs. 3c, 4c and 5c).
Then, one iteratively changes crossings and
unwinds curls until
one gets a trivial knot, keeping track of all moves by the
construction of a polynomial on two variables ($\alpha$ and $z$), one for
crossings and the other for curls,
following prescribed rules \cite{kauf91}.

If $K_1$, $K_2$ and $K_3$ denote the knots displayed in Figs.  3, 4  and 5
respectively, then the polynomials associated with the orbits are

\begin{eqnarray}
H_1(\alpha,z) & = & 1 \\
H_2(\alpha,z) & = & \alpha^{-1} \\
H_3(\alpha,z) & = & \alpha^{-3}. \label{eq:sec}
\end{eqnarray}
\noindent

More important than this particular expression
is the fact that { \em the computation of the invariants gives the same
results for the orbits reconstructed from the observational time series
and for the theoretical orbits. }

In order to reproduce,
in the order of their appearance, the close returns uncovered in the data
equations \ref{eq:prim}
must be integrated with slightly increasing frequencies, as indicated in
the captions of Figs. 3-5.
This is compatible with the structure that shows steadily increasing frequency
displayed in the Gabor transform plot
(Fig. 2).
The last $\frac{1}{3}$ of the observational data contains almost no
dynamics near the $\sim$11 minute period, but does present close returns
near the $\sim$ three minute period.
In equations \ref{eq:prim}, this dynamical behavior is reproduced
when the forcing is eliminated.

We have shown that a low dimensional
dynamical system can account for the pulsations of a rapidly oscillating star.
Moreover, we proposed a model in terms of the normal form of a co-dimension
two bifurcation plus a forcing. Some nonlinear interactions between bifurcating
modes of the basic  equations of stellar dynamics
have been discussed theoretically \cite{buc87}. The equations
describing those interactions are compatible with our
results. This agreement indicates
 that, despite the large number of complex effects taking place, pulsating
stars
can display the nonlinear interaction of a very small number of
 spatially coherent modes.

\begin{center}
{\bf Acknowledgements}
\end{center}

This work has been partially funded by Fundacion Antorchas, Argentina,
and CEE (CI1 CT93-0331).
G. B. Mindlin is a member of CONICET, Argentina. J. L. C. acknowledges
the hospitality of the Observatorio Astron\'omico, UNLP. Arg.
P. T. Boyd acknowledges support from NASA contract NAS5-32490 to USRA.
Additional support was provided by the EUVE program.

\par

\newpage
{\bf{\Large Figure Captions}}
\vspace{7 mm}

Figure 1.   The HST High Speed Photometer observation
of roAp star HD 60435.  Orbital effects have
been eliminated and the data has been rebinned to 30 second sample time.
High frequency components have been eliminated.  For clarity, every
tenth point is plotted.

Figure 2.  Contour plot of the Gabor transform power of the entire observation.
The 11.6 minute period is the fairly flat feature across the bottom of the
plot.  It's strength suddenly diminishes at about 200 minutes into the
observation.  A second feature, with steadily increasing frequency, can be
seen in the upper part of the plot.

Figure 3.  Embedded orbit with HOMFLY polynomial of 1.  In a) the orbit
extracted from the observational data at t=1070 sec is shown.  The numerical
integration
of equations \ref{eq:prim} yield the orbit shown in b), with $\mu=0.028$.
The schematic of the knot, with undercrossings indicated, is displayed in c).

Figure 4.  Embedded orbit with HOMFLY polynomial of $\alpha^{-1}$.
In a) the
observational result at t=2430 sec.
Equations \ref{eq:prim} yield the orbit shown in b), with $\mu=0.030$.
The schematic is displayed in c).

Figure 5.  Embedded orbit with HOMFLY polynomial of $\alpha^{-3}$.
In a) the
observational result at t=13960 sec.
Equations \ref{eq:prim} yield the orbit shown in b), with $\mu=0.033$.
The schematic is displayed in c).

\end{document}